\documentclass[9pt,twocolumn,twoside]{pnas-new}

\templatetype{pnasbriefreport} 

\title{Explaining the low-frequency shear elasticity of confined liquids}

\author[a,c,d]{Alessio Zaccone}
\author[b]{Kostya Trachenko}

\affil[a]{Department of Physics "A. Pontremoli", University of Milan, via Celoria 16,
20133 Milan, Italy.}
\affil[b]{School of Physics and Astronomy, Queen Mary University of London,
Mile End Road, London, E1 4NS, U.K.}
\affil[c]{Department of Chemical Engineering and Biotechnology,
University of Cambridge, Philippa Fawcett Drive, CB30AS Cambridge, U.K.}
\affil[d]{Cavendish Laboratory, University of Cambridge, JJ Thomson
Avenue, CB30HE Cambridge, U.K.}

\leadauthor{Zaccone} 

\authorcontributions{AZ and KT designed research. AZ performed research and calculations. AZ and KT wrote the paper.}
\authordeclaration{There are no competing interests to declare.}
\correspondingauthor{\textsuperscript{2}To whom correspondence should be addressed. E-mail: alessio.zaccone@unimi.it}


\begin{abstract}
Experimental observations of unexpected shear rigidity in confined liquids, on very low frequency scales on the order of 0.01-0.1 Hz, call into question our basic understanding of the elasticity of liquids and have posed a challenge to theoretical models of the liquid state ever since. Here we combine the nonaffine theory of lattice dynamics valid for disordered condensed matter systems with the Frenkel theory of the liquid state. The emerging framework shows that applying confinement to a liquid can effectively suppresses the low frequency modes that are responsible for nonaffine soft mechanical response, thus leading to an effective increase of the liquid shear rigidity. The new theory successfully predicts the scaling law $G'\sim L^{-3}$ for the low-frequency shear modulus of liquids as a function of the confinement length $L$, in agreement with experimental results, and provides the basis for a more general description of the elasticity of liquids across different time and length scales.
\end{abstract}

\dates{This manuscript was compiled on \today}
\doi{\url{www.pnas.org/cgi/doi/10.1073/pnas.XXXXXXXXXX}}

\begin{document}

\maketitle
\thispagestyle{firststyle}
\ifthenelse{\boolean{shortarticle}}{\ifthenelse{\boolean{singlecolumn}}{\abscontentformatted}{\abscontent}}{}


The elasticity of liquids is well understood in the high frequency limit of the mechanical response, where pioneering work by Frenkel~\cite{Frenkel} has shown that the response of a liquid is basically indistinguishable from that of an amorphous solid, provided the frequency of mechanical oscillation is sufficiently high. The idea here is that at short times (high frequency) the diffusive component of the liquid motion is absent and liquids behave as solids. This has become an accepted view \cite{dyre}. However, later experiments have challenged this view \cite{Derjaguin1989,Derjaguin1990,Noirez2006,Noirez2012} and found a remarkable solid-like property of liquids to support shear stress at very low frequency, albeit in confinement. This phenomenon is not currently understood. This is a limitation for the full technological development of small-scale, microfluidic and sub-millimeter flows.

High frequency mechanical response of liquids is typically measured with ultrasonic techniques in the MHz range corresponding to shear elastic moduli of the order of GPa~\cite{Johnson}. The behavior is well described by Frenkel's theory, which links it to transverse acoustic phonons and their vanishing at a characteristic internal time-scale, the Frenkel time, which is related to the viscoelastic Maxwell time.
Conversely, low frequency  shear  elasticity  has been identified fairly recently in view of the long history of liquid research \cite{Trachenko}, starting with the pioneering work of Derjaguin~\cite{Derjaguin1989,Derjaguin1990} and of Noirez~\cite{Noirez2006,Noirez2012} and coworkers. The low-frequency elasticity of liquids is weaker, on the order of $1-10^3$Pa, and  is strongly dependent on the sub-millimeter confinement length-scale of the liquid.

Here we provide a new description of liquid elasticity inspired by Frenkel's ideas on the phonon theory of liquids, combined with recent developments in the microscopic theory of elasticity of amorphous materials. The resulting framework allows us to decompose the various contributions to liquid elasticity based on wavevector $k$, and thus to clearly identify how the shear modulus of a liquid changes upon varying the mesoscale confinement length $L$. 

Following previous literature~\cite{Palyulin}, we introduce the Hessian matrix of the system $H_{ij}=-\partial^2\mathcal{U}/\partial\underline{\mathring{q}}_i\partial\underline{\mathring{q}}_j$ and the affine force field $\underline{\Xi}_{i,\kappa\chi}=\partial\underline{f}_i/\partial\eta_{\kappa\chi}$, where $\eta_{\kappa\chi}$ is the macroscopic strain tensor. Here, $\mathring{q}_i$ is the coordinate of atom $i$ in the reference frame (denoted with the ring notation), whereas $\underline{f}_i=\partial \mathcal{U}/\partial \underline{q}_{i}$ represents the force acting on atom $i$ in the affine position, i.e. in the reference frame subject to a shear (affine) deformation, hence the name "affine" force-field. Greek indices refer to Cartesian components of the macroscopic deformation (i.e. $\kappa\chi = xy$ for shear).

As shown previously, the equation of motion of atom $i$, in mass-rescaled coordinates, can be written~\cite{Lemaitre,Palyulin}:
\begin{equation}
\frac{d^2\underline{x}_i}{dt^2}+\nu\frac{d\underline{x}_i}{dt}+\underline{\underline{H}}_{ij}x_j
=\underline{\Xi}_{i,\kappa\chi}\eta_{\kappa\chi}
\end{equation}
where $\underline{\underline{\eta}}$ is the Green-Saint Venant strain tensor and $\nu$ is a microscopic friction coefficient which arises from long-range dynamical coupling between atoms mediated by anharmonicity of the pair potential. The term on the r.h.s. physically represents the effect of the disordered (non-centrosymmetric) environment leading to nonaffine motions: a net force acts on atom $i$ in the affine position (i.e. the position prescribed by the external strain tensor $\eta_{\kappa\chi}$). As a consequence, in order to maintain mechanical equilibrium on all atoms at each step in the deformation, an additional \textit{nonaffine} displacement is required in order to relax the force $f_{i}$ acting in the affine position, for all atoms. This displacement brings each atom $i$ to a new position which does not coincide with the affine position.

The above equation of motion can be derived from a model particle-bath Hamiltonian as shown in previous work~\cite{Palyulin}. Furthermore, $\{\underline{x}_i(t)=\underline{\mathring{q}}_i(t)-\underline{\mathring{q}}_i\}$, as an expansion around a known equilibrium state $\underline{\mathring{q}}_i$.
Following standard manipulations, which involve Fourier transformation and eigenmode decomposition from time to eigenfrequency~\cite{Lemaitre}, and applying the definition of elastic stress, one obtains the following expression for the
complex elastic constants~\cite{Lemaitre,Palyulin}:
\begin{equation}
C_{\alpha\beta\kappa\chi}(\omega)=C_{\alpha\beta\kappa\chi}^{\textit{Born}}-
\frac{1}{V}\sum_n\frac{\hat{\Xi}_{n,\alpha\beta}\hat{\Xi}_{n,\kappa\chi}}{\omega_{p,n}^2-\omega^2+i\omega\nu} \label{nonaffine}
\end{equation}
where $C_{\alpha\beta\kappa\chi}^{\textit{Born}}$ denotes the affine part of the elastic constant, i.e. what survives in the high-frequency limit. Also, $\omega$ denotes the oscillation frequency of the external strain field, whereas $\omega_p$ denotes the internal eigenfrequency of the liquid (which results, e.g., from diagonalization of the Hessian matrix~\cite{Palyulin}). We use the notation $\omega_{p}$ simply to differentiate the eigenfrequency from the external oscillation frequency $\omega$.

In liquids, a microscopic expression for $G_{\infty} \equiv C_{xyxy}^{\textit{Born}}$ is provided by the well known Zwanzig-Mountain formula~\cite{Zwanzig}, in terms of the pair potential $V(r)$ and of the radial distribution function $g(r)$.
The sum over $n$ in \eqref{nonaffine} runs over all $3N$ degrees of freedom (for a monoatomic liquid with central-force pair interaction). Also, we recognize the typical form of a Green's function in the function over which the sum over $n$ is taken, with an imaginary part given by damping and poles $\omega_{p,n}$ which correspond to the eigenfrequencies of the excitations.

As is usual when dealing with eigenmodes, the sum over $n$ (labelling the eigenmode number) can be replaced with a sum over wavevector $\textbf{k}$, with $\textbf{k}^{2}=k_{x}^{2}+k_{y}^{2}+k_{z}^{2}$, and $k_{x}=\pi n_{x}/L$. We then recall that the numerator of the Green's function, which is given by the eigenfrequency spectrum of the affine force field, can be expressed as $\Gamma(\omega_{p})=\langle\hat{\Xi}_{n,xy}\hat{\Xi}_{n,xy}\rangle_{n\in\{\omega_p,\omega_p+\delta\omega_p\}} \approx A \omega_{p}^{2}$, as proved analytically in ~\cite{Zaccone2011} and numericallly in ~\cite{Palyulin}. This parabolic law holds up to high eigenfrequency as shown numerically in~\cite{Palyulin}.

We thus rewrite \eqref{nonaffine} in terms of a sum over $\textbf{k}$ as follows:
\begin{equation}
G^{*}(\omega)=G_{\infty}-
\frac{A}{V}\sum_{\textbf{k}}\frac{\omega_{p,\textbf{k}}^{2}}{\omega_{p,\textbf{k}}^2-\omega^2+i\omega\nu}
\label{k-nonaffine}
\end{equation}
where $A$ is a numerical prefactor.

In isotropic media, eigenmodes can be divided into longitudinal (L) and transverse (T) modes. Therefore we can split the sum in \eqref{k-nonaffine} into a sum over L modes and a sum over T modes,
\begin{equation}
G^{*}(\omega)=G_{\infty}-
A\sum_{\textbf{k}\lambda}\frac{\omega_{p,\textbf{k}\lambda}^{2}}
{\omega_{p,\textbf{k}\lambda}^{2}-\omega^2+i\omega\nu}
\end{equation}
where $\lambda=L, T$. Furthermore, we introduce continuous variables for the eigenfrequencies $\omega_{p}(k)$, by invoking appropriate dispersion relations $\omega_{p,L}(k)$ and
$\omega_{p,T}(k)$ for L and T modes, respectively (as discussed below). Hence, the discrete sum over eigenstates can be replaced, as is standard in solid-state and statistical physics, with a continuous integral in $k$-space, $\sum_{\textbf{k}}... \rightarrow \frac{V}{(2\pi)^{3}}\int...d^{3}k$:
\begin{align}
G^{*}(\omega)=&G_{\infty}
-B \int_0^{k_{D}}\frac{\omega_{p,L}^{2}(k)}{\omega_{p,L}^2(k)-\omega^2+i\omega \nu}k^{2}dk\\ \nonumber
& - B\int_0^{k_{D}}\frac{\omega_{p,T}^{2}(k)}{\omega_{p,T}^2(k)-\omega^2+i\omega \nu}k^{2}dk ,
\end{align}
the upper limit of the integral is set by the Debye cutoff wavevector $k_{D}$, which, in any condensed matter system (be it solid or liquid), sets the highest frequency of atomic vibration.

One should note that while $k$ is in general not a good quantum number in amorphous materials (as the connection between energy and wavevector is no longer single-valued as it is in crystals where Bloch's theorem holds), it still can be used to provide successful descriptions of the properties of amorphous materials and liquids, including the light scattering spectra of liquids from molecular hydrodynamics~\cite{Hansen}.


We now discuss the dispersion relations for longitudinal and transverse excitations in liquids. For example, for the longitudinal modes, one can resort to the Hubbard-Beeby theory of collective modes in liquids~\cite{Hubbard}, which has been shown to provide a good description of experimental data.
As derived by Hubbard and Beeby, we have the following form for the longitudinal dispersion relation in liquids~\cite{Hubbard}:
\begin{equation}
\omega_{p,L}^{2}(k)=\omega_{E}^{2}\left[1- 3 \frac{\sin k R}{k R}-6 \frac{\cos k R}{(k R)^{2}}+6\frac{\sin k R}{(k R)^{3}} \right] \label{Hubbard}
\end{equation}
where $R$ denotes interparticle separation and $\omega_{E}$ is the Einstein frequency. 
As will be shown below, our final result for the low-frequency limit does not depend on the form of $\omega_{p,L}$. However, for the mathematical completeness of the theory it is important to specify which analytical forms for the dispersion relations can be used to predict the response across the whole frequency domain. 


Differently from gapless longitudinal dispersion relations above and generally from phonon dispersion relations in solids, liquids have the gap in $k$-space in the transverse phonon sector. This follows from the dispersion relation,

\begin{equation}
\omega_{p,T}(k) = \sqrt{c^2k^{2}-\frac{1}{4\tau^{2}}}
\label{kgap}
\end{equation}

\noindent where $\tau$ is liquid relaxation time and $c$ is the transverse speed of sound.

Equation (\ref{kgap}) follows from the Maxwell-Frenkel approach to liquids where the starting point of liquid description includes both elastic and viscous response \cite{Trachenko,baggioli} and implies that transverse modes in liquids propagate above the threshold value $k_g=\frac{1}{2c\tau}$, thus setting the gap in momentum space. Ascertained on the basis of molecular dynamics simulations in liquids \cite{yang}, the $k$-gap operates in a surprising variety of areas, including strongly-coupled plasma, electromagnetic waves, sine-Gordon models, relativistic hydrodynamics and holography \cite{baggioli}. 
At the atomistic level, the Frenkel theory attributes $\tau$ to the average time between molecular rearrangements in the liquid \cite{Frenkel}. In the limit of large $\tau$ or viscosity, Eq. (\ref{kgap}) becomes gapless and solid-like.

In a large system, $k_g$ sets the infrared cutoff in a sum or integral over $k$-points. In a confined system with a characteristic size $L$, the lower integration limit becomes

\begin{equation}
k_{min}=\max \left(k_g,\frac{1}{L}\right)
\label{kmin}
\end{equation}

Then,
\begin{align}
G^{*}(\omega)=&G_{\infty}
-B \int_{\frac{1}{L}}^{k_{D}}\frac{\omega_{p,L}^{2}(k)}{\omega_{p,L}^2(k)-\omega^2+i\omega \nu}k^{2}dk\\ \nonumber
& - B\int_{k_{min}}^{k_{D}}\frac{\omega_{p,T}^{2}(k)}{\omega_{p,T}^2(k)-\omega^2+i\omega \nu}k^{2}dk, \label{integrals}
\end{align}

The lower integration limit for the longitudinal modes in the second term is given by the system size $L$. The lower integration limit for the transverse modes in the third term is given by $k_{min}$ in \eqref{kmin}.

\begin{figure}
    \centering
    \includegraphics[width=0.9\linewidth]{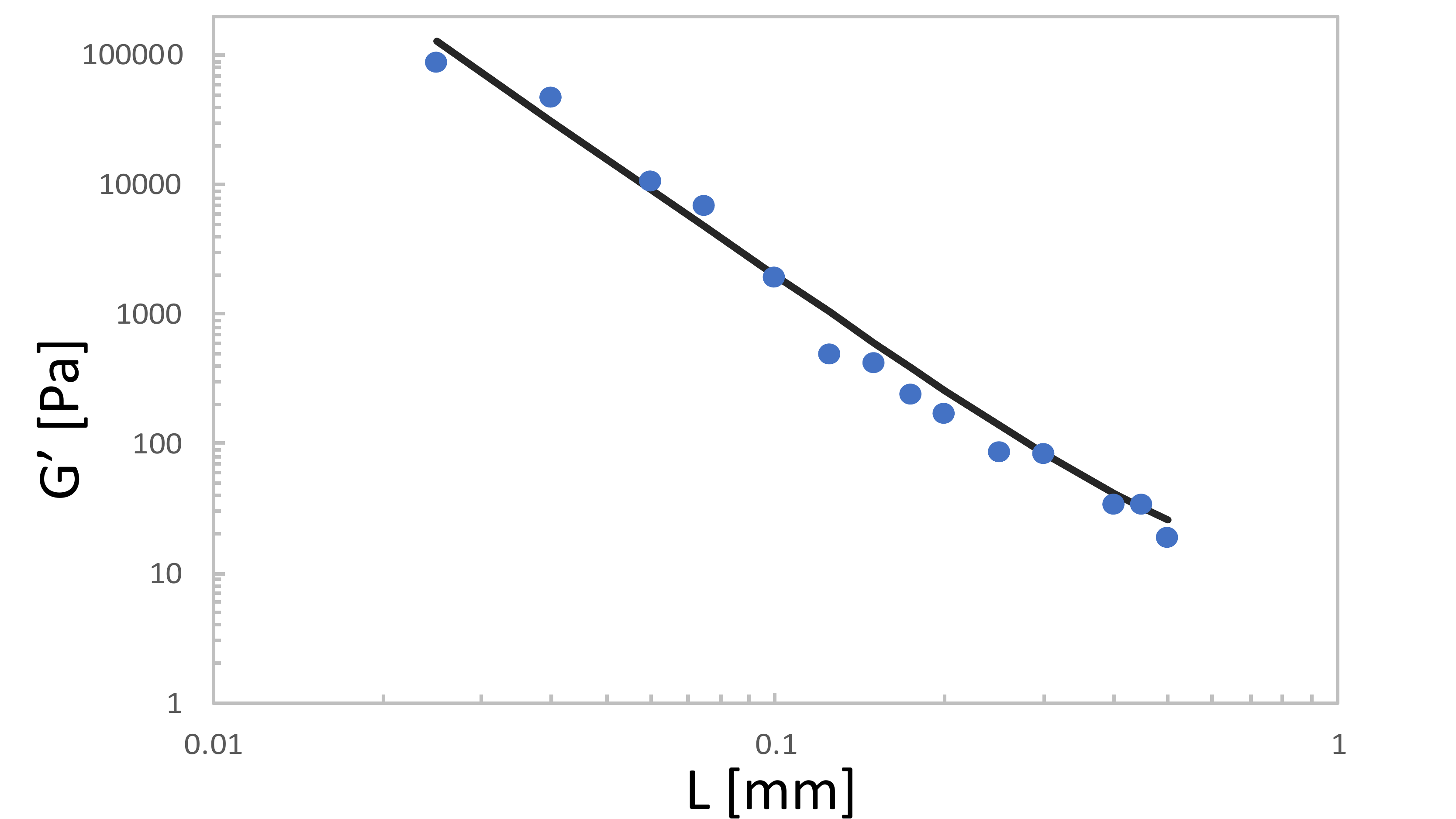}
    \caption{Low-frequency ($\approx 0.01$Hz) storage modulus $G'$ as a function of confinement length $L$. Experimental data refer to short-chain liquid crystalline  polymer liquids PAOCH$_{3}$ (in the isotropic state) well above $T_{g}$~\cite{Noirez2006}, whereas the solid line is the prediction from \eqref{prediction}.}
    \label{fig:uno}
\end{figure}


We take the real part of $G^{*}$ which gives the storage modulus $G'$ and focus on low external oscillation frequencies
$\omega \ll \omega_{p}$ used experimentally. In both integrals numerator and denominator cancel out, leaving the same expression in both integrals. Therefore, as anticipated above, the final low-frequency result does not depend on the form of $\omega_{p,L}(k)$, nor of $\omega_{p,L}(k)$, although the latter, due to the $k$-gap, plays an important role (see \eqref{kmin}) in controlling the infrared cutoff of the transverse integral. In the experiments where the size effect of confinement is seen, $k_g\ll\frac{1}{L}$ ~\cite{Winter}, and $k_{min}=\frac{1}{L}$ according to \eqref{kmin},
leading to
\begin{equation}
G'= G_{\infty} - \alpha \int_{1/L}^{k_{D}}k^{2}dk =G_{\infty} - \frac{\alpha}{3} k_{D}^{3} + \frac{\beta}{3} L^{-3}.   \label{result}
\end{equation}
Here the only term which depends on the system size is the last term, while $\alpha$, $\beta$ are numerical prefactors. In a liquid which is in thermodynamic equilibrium, using the stress-fluctuation version of the nonaffine response formalism (the two versions have been shown to be equivalent in Ref.\cite{Mizuno}) and standard equilibrium statistical thermodynamics, it has been shown by Wittmer and co-workers~\cite{Wittmer} that the affine term $G_{\infty}$ and the negative nonaffine term (here, $- \frac{\alpha}{3} k_{D}^{3}$) cancel each other out exactly, such that $G'(\omega \rightarrow 0) = 0$ for $L \rightarrow \infty$ (bulk liquids). Therefore, for liquids under sub-millimiter confinement, only the third term in the above equation survives, and we obtain
\begin{equation}
G' \approx \beta' L^{-3} \label{prediction}
\end{equation}
where $\beta'=\beta/3$ is a numerical prefactor.

We now compare \eqref{prediction} to available experimental data of low-frequency $G'$ of confined liquids as a function of the confinement length $L$ using the data of the liquid crystalline (LC) short chain polymer in the isotropic state (note that the  formalism of \eqref{nonaffine} is rather general and has been previously successfully tested also for polymer melts in Ref.~\cite{Palyulin}).
In Fig. 1 we compare the trend for storage modulus $G'$ as a function of confinement length $L$ predicted by \eqref{prediction}, with well-controlled experimental data of confined LC-polymer (PAOCH$_{3}$) liquids (in the isotropic state), well above the glass transition temperature $T_{g}$, taken from Ref.~\cite{Noirez2006}. It is evident that the experimental data follow the $L^{-3}$ law predicted in this work. 
Also, in the limit $L \rightarrow \infty$, the above equation \eqref{prediction} recovers the well known result for liquids, i.e. $G'=0$ at low frequency because the third term on the r.h.s. vanishes while the first two terms (affine and nonaffine, respectively) cancel each other out exactly in equilibrium liquids as discussed in~\cite{Wittmer}.

In conclusion, we have developed an analytical theory of the shear modulus of liquids by combining the views of the liquid state as a disordered material originally proposed by Frenkel, with the theory of nonaffine deformation in amorphous solids. This approach allows us to decompose the nonaffine elasticity of the liquid into different phonon-like contributions in terms of their momentum $k$. Since the overall nonaffine contribution to the low-frequency shear modulus is negative (it collects all the relaxational motions which effectively soften the rigidity of an amorphous material), and is expressed as an integral over $k$, the effect of confinement leads to an infrared (long-wavelength) cut-off of the integral which is inversely proportional to confinement size $L$. This explains why reducing the confinement size $L$ effectively increases the shear rigidity by suppressing long-wavelength nonaffine relaxations that soften the response. Ultimately, this framework leads to the law $G' \sim L^{-3}$ for the low-frequency shear modulus of confined liquids in excellent agreement with experimental data of ~\cite{Noirez2006}.
These findings settle a long-standing mystery in our understanding of the liquid state which goes back to the experiments of Derjaguin and, later, of Noirez and co-workers. Furthermore, it may open up new unprecedented avenues for the controlled manipulation of liquids at the micro and nanoscale.

\section*{Acknowledgements}
A.Z. acknowledges financial support from US Army Research Laboratory and US Army Research Office through contract nr. W911NF-19-2-0055. Prof. Laurence Noirez is gratefully acknowledged for providing experimental data in Fig. 1 and for stimulating discussions.


\bibliography{pnas-sample}

\end{document}